\documentclass[12pt,epsf]{iopart}
\usepackage [dvips]{graphicx}
\usepackage{iopams}
\usepackage{epsfig}
\usepackage{verbatim}
\usepackage[usenames,dvipsnames,svgnames]{xcolor}

\addtocounter{secnumdepth}{1}
\font\capfont=cmbx12 at 50 pt 
\newbox\capbox \newcount\capl \def\a{A}
\def\docappar{\medbreak\noindent\setbox\capbox\hbox{%
\capfont\a\hskip0.15em}\hangindent=\wd\capbox%
\capl=\ht\capbox\divide\capl by\baselineskip\advance\capl by1%
\hangafter=-\capl%
\hbox{\vbox to8pt{\hbox to0pt{\hss\box\capbox}\vss}}}
\def\cappar{\afterassignment\docappar\noexpand\let\a }

\begin{document}

\newcommand{\ee}{{\rm e}}
\newcommand{\calT}{{\cal T}}
\newcommand{\calE}{{\cal E}}
\newcommand{\bey}{\boldsymbol{e}_y}
\newcommand{\bv}{\mathbf{v}}

\newcommand{\Jun}{$R_1$}
\newcommand{\Jdeux}{$R_2$}
\newcommand{\Ji}{$R_i$}
\newcommand{\EE}{{E}}
\newcommand{\EEmin}{{\EE}_{min}}
\newcommand{\EEmax}{{\EE}_{max}}
\newcommand{\Ei}{{\epsilon_i}}
\newcommand{\Pexhi}{{P_i^{\rm exh}}}
\newcommand{\Pexhun}{{P_1^{\rm exh}}}
\newcommand{\Pexhdeux}{{P_2^{\rm exh}}}
\newcommand{\gz}{\gamma_0}
\newcommand{\infy}{\inf_{\rule{0mm}{2.55mm}y}}

\newcommand{\bE}{\bar{E}}
\newcommand{\calH}{{\cal H}}
\newcommand{\calJ}{{\cal J}}
\newcommand{\calL}{{\cal L}}
\newcommand{\calM}{{\cal M}}
\newcommand{\calN}{{\cal N}}
\newcommand{\calW}{{\cal W}}
\newcommand{\hP}{\hat{P}}
\newcommand{\hPi}{\hat{\Pi}}
\newcommand{\sumn}{\sum_{n=1}^N}

\newcommand{\Ponewin}{P_{1\rm win}}
\newcommand{\Ptwowin}{P_{2\rm win}}
\newcommand{\Piwin}{P_{i\rm win}}

\newcommand{\vecalpha}{\vec{\alpha}}
\newcommand{\vecg}{\vec{g}}
\newcommand{\vecp}{\vec{p}}

\newcommand{\tA}{\tilde{A}}
\newcommand{\tB}{\tilde{B}}
\newcommand{\tP}{\tilde{P}}
\newcommand{\tbeta}{\tilde{\beta}}
\newcommand{\tgamma}{\tilde{\gamma}}
\newcommand{\tcalM}{\widetilde{\cal M}}
\newcommand{\betast}{{\beta_*}}

\newcommand{\intp}{\int_{-\pi}^{\pi}\frac{\dd p}{2\pi}}
\newcommand{\intpone}{\int_{-\pi}^{\pi}\frac{\dd p_1}{2\pi}}
\newcommand{\intptwo}{\int_{-\pi}^{\pi}\frac{\dd p_2}{2\pi}}
\newcommand{\ointz}{\oint\frac{\dd z}{2\pi{\rm i}}}
\newcommand{\qext}{q_{\rm ext}}

\newcommand{\bO}{{\bf{O}}}
\newcommand{\bR}{{\bf{R}}}
\newcommand{\bS}{{\bf{S}}}
\newcommand{\bT}{\mbox{\bf T}}
\newcommand{\bt}{\mbox{\bf t}}
\newcommand{\half}{\frac{1}{2}}
\newcommand{\thalf}{\tfrac{1}{2}}
\newcommand{\bsA}{\mathbf{A}}
\newcommand{\bsV}{\mathbf{V}}
\newcommand{\bsE}{\mathbf{E}}
\newcommand{\bsT}{\mathbf{T}}
\newcommand{\bsZ}{\hat{\mathbf{Z}}}
\newcommand{\bse}{\mbox{\bf{1}}}

\newcommand{\invup}{\rule{0ex}{2ex}}

\newcommand{\ovG}{\overline{G}}
\newcommand{\fc}{f_{\rm c}}
\newcommand{\gc}{g_{\rm c}}
\newcommand{\tfrac}{\frac}
\newcommand{\xx}{x}
\newcommand{\yy}{y}
\newcommand{\delt}{d}

\newcommand{\dd}{\mbox{d}}
\newcommand{\p}{\partial}

\newcommand{\la}{\langle}
\newcommand{\ra}{\rangle}

\newcommand{\beq}{\begin{equation}}
\newcommand{\eeq}{\end{equation}}
\newcommand{\bea}{\begin{eqnarray}}
\newcommand{\eea}{\end{eqnarray}}
\def\lsim{\:\raisebox{-0.5ex}{$\stackrel{\textstyle<}{\sim}$}\:}
\def\gsim{\:\raisebox{-0.5ex}{$\stackrel{\textstyle>}{\sim}$}\:}

\thispagestyle{empty}



\title{Nash equilibrium in a stochastic model of two competing athletes}
\author{{C\'ecile Appert-Rolland$^{1}$, Hendrik-Jan Hilhorst$^{1}$, and Amandine Aftalion$^{2}$}}

\address{$^1$ Laboratoire de Physique Th\'eorique, Universit\'e Paris-Sud and CNRS UMR 8627 - b\^atiment 210,
91405 Orsay Cedex, France}
\address{$^2$ \'Ecole des Hautes \'Etudes en Sciences Sociales, PSL Research
 University, CNRS UMR 8557, Centre d'Analyse et de Math\'ematique Sociales,
54 Boulevard Raspail, 75006 Paris, France}

\date{\today}

\begin{abstract}


We propose a toy
model for a stochastic description of
the competition between two athletes of unequal strength,
whose average strength difference is represented by a parameter $d$.
The athletes
interact through the choice of their strategies $x,y\in[0,1]$.
These variables denote the amount of energy each invests in the
competition, and determine the performance of each athlete.
Each athlete picks his strategy based on his knowledge
of his own and his competitor's 
performance distribution, and
on his evaluation of the danger of exhaustion,
which increases with the amount of invested energy.
We formulate this problem as a zero-sum game.
Mathematically it is in the class of ``discontinuous games''
for which a Nash equilibrium is not guaranteed in advance.
We demonstrate by explicit construction
that the problem has a mixed strategy Nash equilibrium
$\big( f(x),g(y) \big)$ for arbitrary $0<d<1$.
The probability distributions $f$ and $g$ appear to both be
the sum of a continuous component and a Dirac delta peak.
It is remarkable that this problem is analytically tractable.
The Nash equilibrium provides both the weaker and the stronger athlete
with the best strategy to optimize their chances to win.

\end{abstract}



\section{Introduction}
\label{sec:intro}

The past decade has witnessed a growing use of the tools
of statistical physics in order to understand human systems
involving a form of competition~\cite{chetrite_d_l2017}, for
example in a poker game \cite{sire2007},
but also in sport (e.g. football~\cite{mendes_m_a2007},
tennis~\cite{fischer1980,newton_a2006}, baseball~\cite{sire_r2009},
or sport competitions in general~\cite{bennaim2013}).
A natural framework for describing a competition situation is provided
by game theory. It has been applied, for example, to tournament
competitions~\cite{baek_s_j2015}, or to pedestrians competing for going
through a bottleneck~\cite{bouzat_k2014,heliovaara2011}.

In this paper we focus on competition between two
individual contestants that we will refer to as {\em athletes}.
Our description is general and could apply
to numerous different sports involving
competition between individuals, such as swimming, cycling, rowing, skying, speed-skating. 
For definiteness we will place ourselves in the setting of
running a race.

In the case of running, models at various levels of detail
have been proposed to describe
the supply of energy, both through anaerobic and aerobic metabolisms,
that the athlete will be able to use during the
race~\cite{morton2006}.
This energy supply, together with other parameters such as the
propulsive force or possible friction forces,
will determine a runner’s optimal velocity profile~\cite{keller1974,aftalion_b2014,aftalion2016,aftalion2017}
in the framework of optimal control theory,
with good agreement with real world observations.

In a competition,
interactions of several types may occur
between the runners.
For example, it seems that slipstreaming, well known for cycles, can also
play a role in middle-distance running.
An easy way to model this effect in the frame of optimal control
is to assume that one of the athletes is running as if he were alone,
and to optimize the trajectory of the other runner
under the effect of slipstreaming~\cite{pitcher2009}.
However this approach treats the two athletes on a different footing.
It is possible actually to propose models based on optimal control
for which both trajectories are optimized simultaneously~\cite{aftalion_note}.
In any case, the solution (not necessarily unique) will be deterministic.

In reality the outcome of a race is uncertain due to the
randomness of circumstances beyond the runners' control.
In the present paper we therefore introduce a stochastic model for a
two-runner race, in which stochasticity can be seen as
a modeling of the level of fatigue or motivation that can
vary from one race to the next.
We aim at presenting an exactly solvable model
that illustrates principles, rather than retaining full realism.
In our toy model,
both runners are treated on equal footing.
In particular, both are free to mutually adapt their strategies, 
and each of them does so with the purpose
of winning as many races as possible.
The stochasticity in the description opens
in particular
the possibility for the weaker athlete to win 
{\it with a certain probability}. 
The model will take the form of a game
and we will therefore be able to
apply the tools of game theory,
whose essential points we  will recall as we go along.

Assuming that the same two runners compete through a large number
of races, we will find analytically a Nash equilibrium,
that is a set of strategies such that none of the runners
can increase his gain by changing unilaterally his strategy.
We will show that in order to reach this Nash equilibrium, runners should not
run  all races the same way, but rather use mixed strategies so as to surprise
their competitor.

In section \ref{sec:model} we define the model and express it in a form
amenable to a game theoretic solution. 
In section \ref{sec:nonexistence} we show that if the model has a Nash
equilibrium, this equilibrium consists necessarily of mixed strategies.
In section \ref{secsymm} we briefly recall the symmetric case,
well-known in the literature,
in which the athletes are of equal strength. 
In section \ref{secasymm} we derive the main results of our work.
We consider a weaker and a stronger athlete and show how
this changes the nature of the symmetric solution.
The full asymmetric solution is calculated analytically and commented upon.
In section \ref{sec:conclusion} we conclude and
discuss some future perspectives.


\section{A two-athlete model}
\label{sec:model}

\subsection{Running times and energies}

We consider two runners \Jun\ and \Jdeux.
For each race, we associate with \Ji\
a variable $\EE_i$\, $(i=1,2)$, that we will refer to as his 
{\em energy.}
When runner \Ji\ has an energy $\EE_i$, he will complete the race 
within his best
final time $T(\EE_i)$,
which we stipulate to be a decreasing function of its argument.

The introduction of this ``energy''variable requires a short discussion.
The runner's best final time depends on several parameters, 
including various types of energy supplies (anaerobic, VO2%
\footnote{The VO2 is the rate at which oxygen is transformed into energy.}) 
but also other physiological characteristics 
such as his maximal propulsive force
or his friction coefficient.
We refer to \cite{aftalion2016,aftalion_b2014,aftalion2017} for an analysis of
how the final time depends on these parameters individually.
The ``energy'' $\EE_i$\  should be thought of as describing
the integrated effect of all these parameters, in the spirit of
the {\em invested stamina} introduced in~\cite{baek_s_j2015}.
Actually the variable $\EE_i$\ could also simply be thought of
as a way to parametrize the distribution of final times.

For each race, the energy $\EE_i$
available to runner \Ji\
is taken to be an independent stochastic variable
drawn from a probability distribution $\rho_i(\EE_i)$. 
Each runner has a knowledge of the two energy
distributions $\rho_i$ ($i=1,2)$ and thus
knows in particular whether {\it on average\,}
he is stronger or weaker than his opponent.
But on a given race, he can only make a guess
about the {\it actual\,} random values $\EE_i$ ($i=1,2)$ of the
energies that are going to be available
to himself and his opponent.
Runner $R_i$ will therefore choose a value $\Ei$ (his ``strategy'')
as a guess for the energy he expects to have
available for this particular race.

Two scenarios are then possible.
Either \Ji\ overestimated the energy available to himself
($\Ei > \EE_i$) and he will
be exhausted before reaching the finish line;
or he underestimated it ($\Ei \le \EE_i$)
and he will finish the race in a time $T(\Ei)$.
The probability of exhaustion associated with the choice 
of strategy $\Ei$ is thus
\beq
\Pexhi = \int_{-\infty}^\Ei \rho_i(\EE_i)\, \dd\EE_i
\equiv \left\{
\begin{array}{rr}
x, \qquad & i=1,\\[2mm]
y, \qquad & i=2.
\end{array}
\right.
\label{eq:pexh}
\eeq

It is now clear how the interaction between runners comes into play.
If a runner knows that he runs against a stronger competitor,
he will tend to choose a higher $\Ei$ in order to realize
a shorter time  $T(\Ei)$, in spite of the increased risk of
failing by exhaustion.
If he knows he runs against a weaker opponent,
he will choose a lower $\Ei$
in order to reduce his risk of exhaustion.

Our term ``exhaustion'' should be understood in a broad sense.
Indeed, in the same way as the variable $\EE_i$ represents the aggregation
of several parameters, ``exhaustion'' also covers a variety of situations.
For example, a lack of concentration can lead to a false start,
athletes may suffer from various injuries, etc.
Besides, exhaustion does not necessarily mean that the runner does not
reach the finish line, but that he would do it (if at all) with a final time
far less than the best time he could have expected
(at least less than any best time $T(\EE_i)$ that the other runner can achieve).
There are (rare) examples in the sport history
where a weaker competitor indeed won a competition because other ones
were eliminated for various reasons. For example, in the Olympic Games
of Sydney in 2000, one Guinean swimmer won a race though it was the
first time he was swimming in a 50 meter swimming pool,
just because the other ones both made false starts~\cite{moussambani2000}.

We finally observe that this simplified model does not take into account any
interactions between the two runners that might occur {\it during\,} the race.
Such more difficult questions are left for follow-up work.


\subsection{Choice of the energy distributions $\rho_i(\EE_i)$}

\begin{figure}[t]
\begin{center}
\scalebox{.55}
{\includegraphics{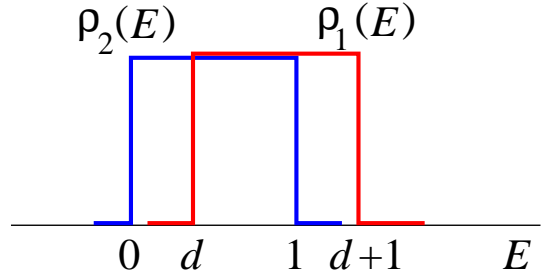}}    
\end{center}
\caption{\small The energy distributions $\rho_1(E)$ and $\rho_2(E)$
of runners \Jun\ and \Jdeux.}
\label{fig:energyscale}
\end{figure}

To simplify the calculations we will take for each $\rho_i(\EE_i)$
a uniform energy distribution on a unit interval.
This interval will be $d\leq\EE_1\leq d+1$ for runner \Jun\
and $0\leq\EE_2\leq 1$ for runner \Jdeux, as shown in Figure
\ref{fig:energyscale}.
For $d=0$ there is symmetry between the two runners; for $d>0$
runner \Jun\ is the stronger one.
Except when stated otherwise, we will consider only the case $0\le d<1$,
for which the weaker runner still has a nonvanishing probability to win.
The case $d\ge 1$ is trivial in the sense that the stronger runner will always win.

For each race a stochastic variable
$\EE_i$ (unknown to the runners) is chosen
for each runner according to its distribution $\rho_i$.
When we apply (\ref{eq:pexh})
for given guesses $\epsilon_i$ of the energies,
we find that\footnote{{Here exhaustion would mean that either the runner
does not reach the finish line, or that it does so within a running time
which is larger that the largest $T(E)$ of the other runner,
i.e. $T(\epsilon=0)$ or $T(\epsilon=d)$.}}
\beq
\Pexhun = x = \epsilon_1 - d, \qquad \Pexhdeux = y = \epsilon_2\,.
\label{defxy}
\eeq
Hence the choice of the strategy pair $(\epsilon_1,\epsilon_2)$
is equivalent to the choice of the pair $(x,y)$;
since both $x$ and $y$ vary in the unit interval, these will be the most
convenient variables for the calculations in later sections.


\subsection{The resulting game}
\label{sec:game}

For a given strategy pair  $(\epsilon_1,\epsilon_2)$
the possible outcomes of the race are listed in Table \ref{table1} together
with their probabilities of occurrence.
Four cases (first column of the table) must be distinguished,
according to whether
none, only \Jun, only \Jdeux, or both runners are exhausted
before reaching the finish line.
These cases occur with the probabilities listed in the second column of the
table, and the winner is given by the third column.
Indeed,
when both runners reach the finish line without being exhausted,
the winner is the one with the shorter time $T(\Ei)$, hence the larger
energy $\Ei$; in case $\epsilon_1=\epsilon_2$,
which is most of the time of measure zero
as we shall see later,
the winner is chosen randomly
with probabilities $\tfrac{1}{2}(1\pm\gz)$
where,
we introduced $\gz\in[-1,1]$ as an additional
parameter of the problem
for reasons that will become clear later.
When only one of the runners is exhausted, the other one wins.
When both are exhausted, the winner is chosen uniformly randomly.
There thus is always a winner and a loser.
\begin{table}
\begin{center}
{\small
\begin{tabular}{|l|l|l|}
\hline
\mbox{Exhausted runners} & Probability & Winner\\
\hline
\mbox{none} & (1-$\Pexhun)(1-\Pexhdeux$)
  & \mbox{if } $\epsilon_1>\epsilon_2$~:
    \mbox{ \Jun} \\
& & \mbox{if } $\epsilon_2>\epsilon_1$~:
    \mbox{  \Jdeux} \\
& & \mbox{if } $\epsilon_1=\epsilon_2$~:
    \mbox{ \Jun\ with probability} $\tfrac{1}{2}(1+\gz)$\\
& & \phantom{\mbox{if } $\epsilon_1=\epsilon_2$~:}
    \mbox{ \Jdeux\ with probability} $\tfrac{1}{2}(1-\gz)$\\
\mbox{\Jdeux\ only} & $(1-\Pexhun)\Pexhdeux$ & \mbox{\Jun} \\
\mbox{\Jun\   only} & $\Pexhun(1-\Pexhdeux)$ & \mbox{\Jdeux} \\
\mbox{both} & $\Pexhun \Pexhdeux$
& \mbox{\Jun\,, \Jdeux\ with equal probability} \\
\hline
\end{tabular}
}
\caption{Race winner shown in four different cases of exhaustion
and relative values of the $\epsilon_i$.
The physical variables $\Pexhun, \Pexhdeux, \epsilon_1$, and $\epsilon_2$
may all be expressed in terms of $x$ and $y$ with the aid of
Eq.\,(\ref{defxy}).}
\label{table1}
\end{center}
\end{table}
We now model this race as a game by assigning a payoff $+1$ to the winner and
$-1$ to the loser,
which means that we have a zero-sum game.
For a given strategy pair
$(\epsilon_1,\epsilon_2)$  the {\it expected\,}
payoff for \Jun, to be denoted as $G_1(\epsilon_1-d,\epsilon_2)=G_1(x,y)$,
is obtained as the sum on the probabilities for
the four cases listed in Table 1, each weighted with
$\pm 1$ according to whether runner \Jun\ is or is not the winner.
Upon using Eq.\,(\ref{defxy}) to express
all quantities in terms of the variables $x$ and $y$ we find
\begin{figure}[hbt]
\begin{center}
\scalebox{.45}
{\includegraphics{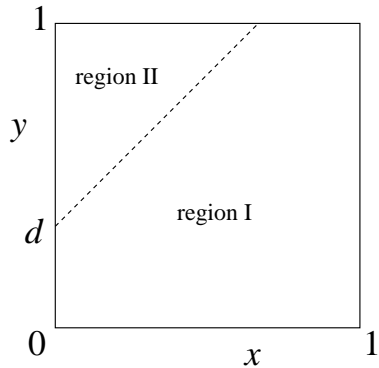}}    
\end{center}
\caption{\small The payoff function $G_1(x,y)$, defined on the unit square,
is discontinuous along the dashed line separating
regions I and II.}
\label{figsquare}
\end{figure}
\beq
G_1(x,y) = \left\{
\begin{array}{l l}
-1+2y-xy, \qquad &  x+\delt<y \quad \mbox{ (region II),}\\[2mm]
\phantom{-}1-2x+xy, \qquad & x+\delt > y \quad \mbox{ (region I),}\\[2mm]
\phantom{-}d + \gamma(x), \qquad & x+d = y,
\end{array}  \right.
\label{game_asym}
\eeq
with $(x,y) \in [0,1]^2$ and in which
\beq
\gamma(x) =\gz\,(1-x)(1-x-d)
\label{dgamma}
\eeq
defines the payoff on the line of discontinuity.
For $\gz=0$ this payoff
takes the  value $d$, halfway between the neighboring values
in the regions I and II (see Figure \ref{figsquare}).
For $\gz=1$ [for $\gz=-1$] the value $d+\gamma(x)$
is equal to the limit of $G_1(x,y)$ as the discontinuity is approached
in region I [in region II], so that
$G_1$ is upper (lower) semi-continuous.

Since the race is a zero-sum game,
the expected gain $G_2(y,x)$
of runner \Jdeux\ is
\beq
G_2(y,x) = - G_1(x,y).
\label{eq_G2}
\eeq
Eq.\,(\ref{game_asym}) together with (\ref{eq_G2}) defines the ``game.''

We will now address the question of determining
the best strategies for $R_1$ and $R_2$
if each wants to maximize his payoff.
This is the subject of the next sections.


\section{Nonexistence of a Nash equilibrium of pure strategies}
\label{sec:nonexistence}

Having defined the game (\ref{game_asym}),
we now ask what the best strategy is for each runner
and will appeal to game theory to find the answer.

In general, a strategy will be {\em mixed,} that is,
each time the race is repeated,
runner \Jun\ chooses an $x$ from
a distribution $f(x)$ and runner \Jdeux\ a $y$ from a distribution $g(y)$.
A {\em pure\,} strategy is the special case
in which a runner always chooses the same $x=x_0$ or $y=y_0$
(hence $f$ or $g$ is a Dirac delta function).
A Nash equilibrium is a pair of strategies $(f(x),g(y))$
such that neither runner
can improve his payoff by a unilateral change of strategy.
A general game may have no such equilibrium, or a unique one, or several.

We will now show that (\ref{game_asym}) does not have a Nash equilibrium
of pure strategies.
For a zero-sum game, game theory tells us that a
Nash equilibrium
coincides with a minimax solution.
The idea of a minimax solution is that each player will try to minimize
his worst-case loss, or equivalently that he will try to maximize
his worst-case gain.
When we restrict ourselves to pure strategies,
this means that runner \Jun\ will play the strategy $x^*$
that ensures he will win at least
$G^*_1 = \max_{x} \left[\min_{y} G_1(x,y) \right]$
and runner \Jdeux\ will play the strategy $y^*$
that ensures he will lose at most
$-G^*_2 = \min_{y} \left[\max_{x} G_1(x,y) \right]$.
If $G_1^*$ and $G_2^*$ exist, then necessarily
\begin{equation}
G^*_1  =  \max_{x} \min_{y} G_1(x,y) \le
\min_{y} \max_{x} G_1(x,y) = -G^*_2
\label{minimaxineq}
\end{equation}
If, moreover, for a pair $(x^*,y^*)$ of strategies Eq.\,(\ref{minimaxineq})
holds as an equality,
then this pair is a minimax solution and
is also a Nash equilibrium.
Inversely, if there exists a pure strategy Nash equilibrium,
then $G_1^*$ and $G_2^*$ should both exist and be equal.

For our present problem, and for $0\le d < 1$, one shows  with the aid of a short calculation that
\bea
\sup_{x} \infy G_1(x,y) &=& d - \Delta,
\noindent\\[2mm]
\infy \sup_{x} G_1(x,y)
&=& d + \Delta, \qquad \Delta\equiv 3-\sqrt{8+d^2},
\label{xsupinf}
\eea
which excludes the possibility for
Eq.\,(\ref{minimaxineq}) to be satisfied with the equality sign.
Hence we see that there cannot be a Nash equilibrium of pure
strategies\footnote{In the case $d\ge 1$, the stronger runner
will always win. Then a trivial Nash equilibrium is obtained
when the stronger runner plays the pure strategy
$f(\xx) = \delta(\xx)$. Any strategy of the weaker runner will
lead to the same payoff $G_2=-G_1=-1$. Thus an infinity of strategies $g(y)$
can be chosen which all lead to a Nash equilibrium.}.
We have relegated the proof of (\ref{xsupinf}) to \ref{sec:AppendixA},
where we show that the extrema are reached at the point $(x^*,x^*+d)$
with $x^*=(4-d-\sqrt{8+d^2})/2$, located on the line of discontinuity,
and that the difference $2\Delta$ between the two expressions in
(\ref{xsupinf}) is equal to the jump of $G_1$ across the discontinuity in that
point.

In the absence of a pure strategy Nash equilibrium,
the next step is to look for one with mixed strategies.
The fact that in our game there is a discontinuity will
turn out to be mathematically essential and we will therefore
treat the line $x+d=y$ discontinuity with the care required.
Glicksberg's theorem (as cited by~\cite{dasgupta_m1986}) guarantees
the existence of a Nash equilibrium for a discontinuous game
only if $G_1(x,y)$ is upper or lower semicontinuous.
The game (\ref{game_asym}) lacks this property for all $\gz\in(-1,1)$,
but we will show by explicit construction
that it nevertheless does have a Nash equilibrium.


\section{Symmetric problem}
\label{secsymm}

We take in this section $d=0$, so that we have
two runners of equal strength,
\Jun\ and \Jdeux, with respective strategies $x,y \in [0,1]$.
The (positive or negative) gain of \Jun\ is given by the payoff function
\beq
G_1(x,y) = \left\{
\begin{array}{ll}
-1+2y-xy, \qquad & x<y,\\[2mm]
\phantom{-}1-2x+xy,  & x>y, \\[2mm]
\phantom{-}\gz(1-x)^2,  &  x=y,
\end{array}
\right.
\label{payoffS}
\eeq
and the gain of \Jdeux\ is the opposite, $G_2(y,x)=-G_1(x,y)$.
This game (with $\gz=0$) is a classical textbook example~\cite{dresher1961}.
We briefly recall how its Nash equilibrium is obtained.
Since we found in section \ref{sec:nonexistence} that this problem has
no Nash equilibrium of pure strategies,
we will now look for one in which
\Jun\ and \Jdeux\ have mixed strategies with identical
distributions $f(x)$ and $f(y)$,
respectively, that satisfy
\beq
f(z) \geq 0, \qquad \int_0^1 {\rm d}z\,f(z) =1.
\label{normcond}
\eeq
The support of $f(z)$ may be smaller than the full interval $[0,1]$.
One may try to solve the problem by supposing
that the support is an interval
$[0,a]$ with an as yet unknown $a\in(0,1)$.

\noindent
If runner \Jun\ chooses a strategy $x$ in the support of $f(x)$,
then his expected payoff averaged over $R_2$'s mixed strategy $f(y)$ will be
\beq
\ovG_1[x;f(y)] = \int_0^a {\rm d}y\, G_1(x,y)f(y).
\label{defG1}
\eeq

As a consequence of the definition of a Nash equilibrium,
the mixed strategy $f(y)$ that we are looking for
should be such that
\beq
\ovG_1[x;f(y)] = K, \qquad 0\leq x\leq a,
\label{eqzero}
\eeq
where $K$ is some constant.
Indeed, if a specific strategy $x=x_0$ would yield a payoff
less than $K$ to runner
$R_1$, then $R_1$ would remove that $x_0$ from
the support of $f(x)$; and if a specific strategy $x=x_0$ would yield a
payoff larger than $K$ to runner $R_1$, then $R_1$ could improve his expected
payoff by
putting a larger weight on that $x_0$
-- both in contradiction
with the definition
of a Nash equilibrium.
By symmetry we must of course in the end find $K=0$,
but we have not needed that here.
\vspace{2mm}

\begin{figure}[t]
\begin{center}
\scalebox{.55}
{\includegraphics{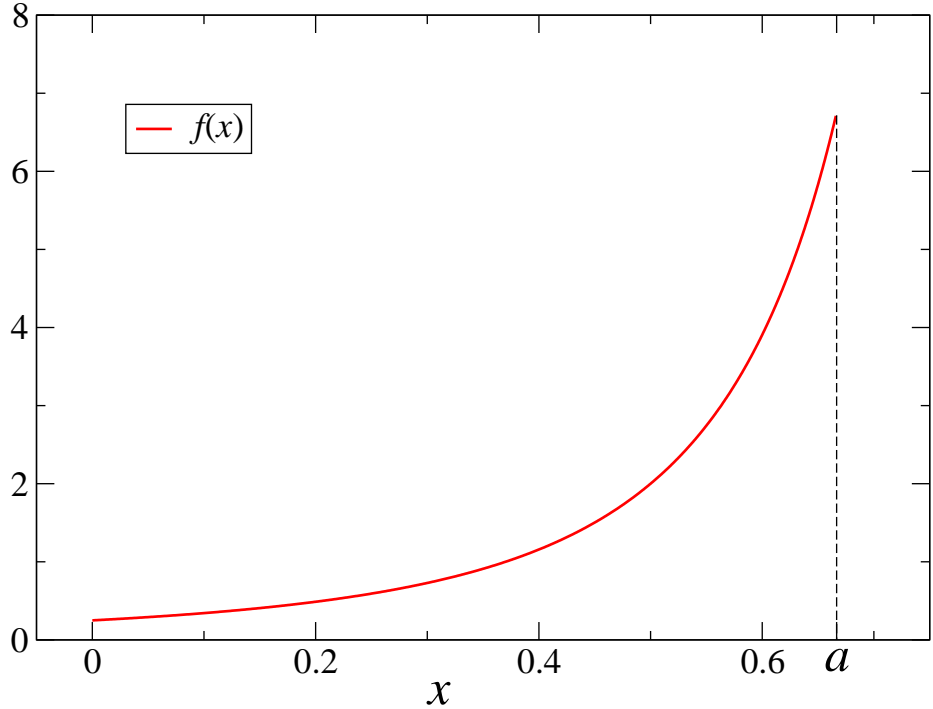}}    
\end{center}
\caption{\small The function $f(x)$ representing the Nash equilibrium
for the symmetric problem, $d=0$. Here $a=\tfrac{2}{3}$. }
\label{fig:fx}
\end{figure}

It is not {\it a priori\,} clear how many solutions there are to
(\ref{eqzero}), if any at all.
We will look for an $f$ that is differentiable.
When we substitute (\ref{defG1}) in (\ref{eqzero}),
use the explicit expression (\ref{payoffS}) for the payoff function, and
differentiate twice with respect to $x$, there arises
a first order ODE for $f(y)$ whose solution is
\beq
f(y) = \frac{D(a)}{(1-y)^3}\,, \qquad 0\leq y \leq a.
\label{solutionf}
\eeq
Normalization yields $D$ as a function of the
hypothesized interval parameter
$a$,
\beq
D(a) = \frac{2(1-a)^2}{a(2-a)}\,.
\label{relDa0}
\eeq
In order to determine $a$ we substitute (\ref{solutionf})
in the original expression
(\ref{eqzero}), which leads to
\beq
K = \frac{(3a-2)D}{2(1-a)^2} \left[a - (2-a)x \right],
\qquad 0\leq x\leq a.
\label{xKS}
\eeq
We should now render the RHS of (\ref{xKS}) independent of $x$.
As $a$ cannot be equal to $2$ (we have $a<1$), the only
way this condition can be met is when
the prefactor $3a-2$ vanishes. Then $K=0$ as expected and
$a=\frac{2}{3}$, which combined with (\ref{relDa0}) yields
$D=\frac{1}{4}$.
The resulting function $f(x) = 1/[4(1-x)^3]$ is shown in
Figure \ref{fig:fx}.
To make sure that it actually represents a Nash equilibrium, one also has to show
that neither runner can improve his result by choosing a strategy outside of
the support $[0,a]$. This is easily done \cite{dresher1961}.
Finally we observe that the value of $G_1(x,y)$ on the
discontinuity line has played no role here: the equilibrium is
independent of $\gz$.

\section{Asymmetric problem}
\label{secasymm}

The case of asymmetric runners has not so far been studied,
and doing so is the purpose of this work.
The question of whether a discontinuous game has or does not have
a Nash equilibrium
is the subject of several mathematical theorems,
but has not received an exhaustive answer.
Glicksberg's theorem applies here only for $\gz = \pm1$.
We don't know either whether the solution is unique.
We will show below that for the problem at hand
such a solution does exist by calculating it explicitly.
It is remarkable that this problem may be solved analytically
exactly, as there is no general method allowing to systematically
find Nash equilibria.


\subsection{The asymmetric problem and the {\it ansatz\,} for its  solution}

For an asymmetry parameter $d>0$ defined in section \ref{sec:game}
the payoff function $G_1(x,y)$ that gives runner $R_1$'s gain is given by
the general expression Eq.\,(\ref{game_asym}), which we repeat here for easy
reference,
\beq
G_1(x,y) = \left\{
\begin{array}{ll}
-1+2y-xy,                 \qquad & x+d<y,\\[2mm]
\phantom{-}1-2x+xy,       \qquad & x+d>y, \\[2mm]
\phantom{-}d + \gamma(x), \qquad & x+d=y.
\end{array}
\right.
\label{payoff}
\eeq
The gain of runner \Jdeux\ is again equal
to $R_1$'s loss, that is, $G_2(\yy,\xx)=-G_1(\xx,\yy)$.
We note that the payoff function (\ref{payoff})
differs from the one of the symmetric case,
Eq.\,(\ref{payoffS}),
only by a parallel shift of the line of discontinuity.
In this case we have to solve the full asymmetric game, that is, assume that
runners \Jun\ and \Jdeux\
have distinct strategies $f(\xx)$ and $g(\yy)$, respectively.
Their expected gains $\ovG_1[\xx;g(\yy)]$ and
$\ovG_2[\yy;f(\xx)]$ have the expressions
\bea
\ovG_1[\xx;g(\yy)] &=& \int_{{\rm supp}\,g} {\rm d}\yy\, G_1(\xx,\yy)g(\yy),
\label{xES1}\\[2mm]
\ovG_2[\yy;f(\xx)] &=& \int_{{\rm supp}\,f} {\rm d}\xx\, G_2(\xx,\yy)f(\xx).
\label{xES2}
\eea
and we must consider the pair of equations
\bea
\ovG_1[\xx;g(\yy)] = \ \ K, \qquad && \xx\in\mbox{supp } f,
\label{eqzerox}\\[2mm]
\ovG_2[\yy;f(\xx)] = -K,    \qquad && \yy\in\mbox{supp } g.
\label{eqzeroy}
\eea
As \Jun\ is now stronger than \Jdeux,
we anticipate that $K>0$.

There is no general method to determine $f$ and $g$.
In the preceding section we assumed that the solution
$f=g$ was
differentiable, and it appeared that indeed such a solution existed;
but this need not always be the case.
In the asymmetric case two arguments can help us to guess
the forms of $f$ and $g$.
First, the solution to the symmetric problem given above suggests
that $f$ and $g$ will at least each have
a differentiable {\it component}, in addition to anything else.
Secondly, we have used the online solver of
Avis {\it et al.}\,\cite{avis2010}
to obtain the exact Nash equilibrium
for a version of the system
in which each strategy was confined to a set of fifteen discrete points,
with the points of the two sets alternating along the axis.

\begin{figure}[hbt]
\begin{center}
\scalebox{.45}
{\includegraphics{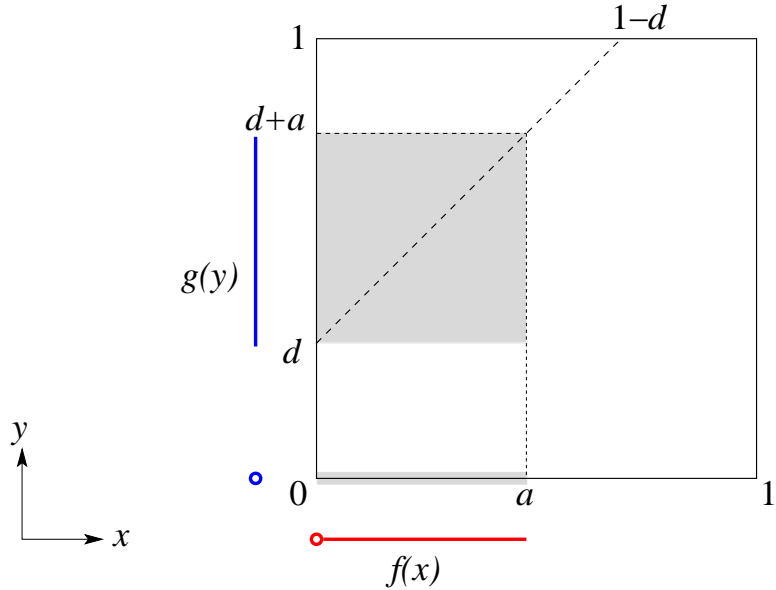}}
\end{center}
\caption{\small The space of possible strategies is the unit square.
The payoff function $G_1(x,y)$ is discontinuous along the dashed
line parallel to the diagonal.
The Nash equilibrium is a pair $\big(f(x),g(x)\big)$.
The heavy red line, when projected vertically
onto the $x$ axis, gives the support of
$f(x)$; similarly, the heavy blue one, when projected horizontally
onto the $y$ axis, gives the support of $g(y)$.
The red and blue circles denote the location of Dirac
delta function components. The one of $g(y)$ is isolated from the rest of its
support.
The effectively used strategy pairs $(x,y)$ in the Nash equilibrium
are limited, therefore, to the subspace
$[0,a] \times \Big( \{0\}\cup(d,d+a] \Big)$ that has been shaded and
consists of a square and a line interval.}
\label{figxysquarelarge}
\end{figure}

Inspired by these considerations we decided to look for
strategies $f(\xx)$ and $g(\yy)$ that both have a differentiable
component and, at the
lower end of the allowed strategy interval, a Dirac delta peak.
That is, we make the {\it ansatz}
\bea
f(\xx) &=& q_1\delta(\xx)   + (1-q_1)\fc(\xx),  \qquad
0\leq\xx\leq 1,
\label{hypf}\\[2mm]
g(\yy) &=& q_2\delta(\yy) + (1-q_2)\gc(\yy),  \qquad
0\leq\yy\leq 1,
\label{hypg}
\eea
where $\fc$ and $\gc$ are differentiable and
have their support on an interval $\epsilon_1,\epsilon_2\in(d,d+a]$,
that is common to both runners on the original energy axis.
Here again, the interval length $a$ is as yet unknown
but should be such that $0<a\leq 1-d$.
In terms of the variables $x$ and $y$ this
``common'' interval is given by
\beq
0 < \xx\leq a, \qquad  d < \yy\leq d+a.
\eeq
Hence
\beq
\fc(\xx),\ \gc(\yy) > 0, \qquad \int_0^{a}\dd\xx\,\fc(\xx)
=\int_d^{d+a}\dd\yy\,\gc(\yy) = 1.
\label{normcond2}
\eeq
A schematic representation of the space of possible strategies
is given in Figure \ref{figxysquarelarge}.


\subsection{Determining $\fc(\xx)$, $\gc(\yy)$,  $q_1$,  $q_2$, and the
  interval length $a$}

We proceed
to find a pair of strategies $f(x)$ and $g(y)$
having the hypothesized form (\ref{hypf})-(\ref{hypg}).

Upon substituting  Eqs.\,(\ref{hypf}) and (\ref{hypg})
in Eqs.\,(\ref{xES2}) and (\ref{xES1}), respectively,
we obtain
\bea
\ovG_1[x;g(y)] &=&
q_2 G_1(x,0) + (1-q_2)\int_d^{d+a}\dd y\,G_1(x,y)g_{\rm c}(y),
\nonumber\\
& & 0\leq x\leq a
\label{eqGxg}
\eea
and, using the zero-sum property (\ref{eq_G2}) to eliminate $G_2$ in favor of
$G_1$\,,
\bea
\ovG_2[y;f(x)] &=&
-q_1 G_1(0,y) - (1-q_1)\int_0^a\dd x\,G_1(x,y)f_{\rm c}(x),
\nonumber\\[2mm]
& &y=0 \ \ \mbox{ or } \ \ d < y\leq d+a.
\label{eqGyf}
\eea
On the RHS of these
equations the contribution of the Dirac deltas are
respectively
$G_1(x,0)=1$ and $G_1(0,y)=-1+2y$.
In the integrals appearing in Eqs. (\ref{eqGxg})-(\ref{eqGyf}),
we now use
the explicit expression (\ref{payoff}) for $G_1(x,y)$.
This leads to splitting both intervals of integration
into two subintervals
separated by a point of discontinuity at $x+d=y$.
The value $G_1(x,x+d)=d+\gamma(x)$ of the payoff {\it at\,} the discontinuity
has zero weight under the integrals and does not affect the outcome.
Note that in
Eq.\,(\ref{eqGyf}) the particular value
$y=d$, which is outside the support of $g(y)$, is excluded.
As a consequence the value of the payoff on the line of
discontinuity has not so far appeared in any of our considerations.

\begin{figure}[hbt]
\begin{center}
\scalebox{.45}
{\includegraphics{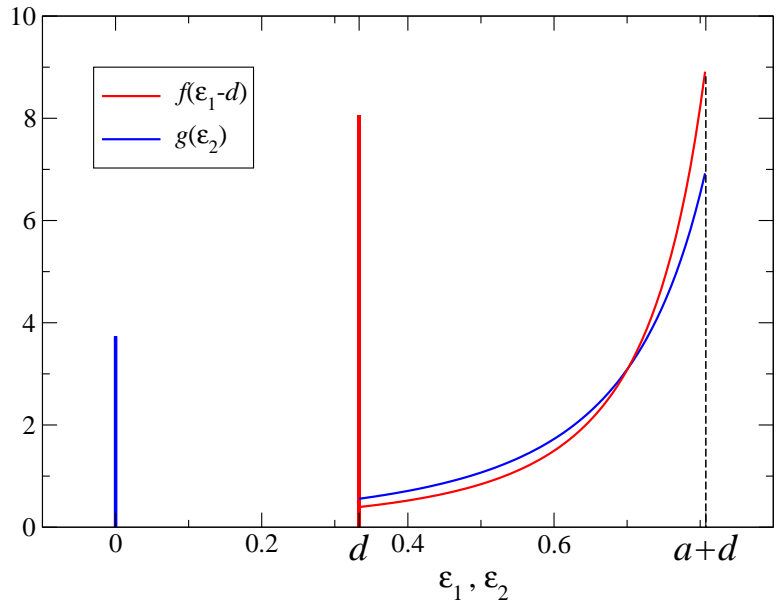}}
\end{center}
\caption{\small
The two strategies for asymmetry parameter $d=\tfrac{1}{3}$
as functions of the energies $\epsilon_1=\xx+d$ and $\epsilon_2=\yy$.
Each vertical bars represents the value of the integrated delta function
at that point, multiplied by 100. The continuous components
of the strategies are
$\fc(\xx)$ (red curve) and $\gc(\yy)$ (blue curve). The right hand limit point
of the support is at $a+d
=0.4743799+\tfrac{1}{3}
=0.8077132$.}
\label{fig4}
\end{figure}

When differentiating (\ref{eqGxg}) twice with respect to
$\xx$ and (\ref{eqGyf}) twice with respect to $\yy$,
we find  for $\fc(\xx)$ and $\gc(\yy)$ the linear first order ODEs
\beq
f^\prime_{\rm c}(x) = \frac{3-d-3x}{(1-x)(1-d-x)}f_{\rm c}(x), \quad
g^\prime_{\rm c}(y) = \frac{3+d-3y}{(1-y)(1+d-y)}g_{\rm c}(y),
\label{odefcgc}
\eeq
in which the parameters $q_1$, $q_2$, and $K$
no longer appear.
The solutions are
\bea
\fc(\xx) &=& \frac{B(a)}{(1-d-\xx)^2(1-\xx)}\,, \qquad 0 < x \leq a,
\label{solutionfc} \\
\gc(\yy) &=& \frac{D(a)}{(1+d-\yy)^2(1-\yy)}\,, \qquad d < y \leq d+a,
\label{solutiongc}
\eea
and vanish outside the intervals indicated.
Normalization yields $D$ and $B$ in terms of the interval
length $a$,
\bea
D^{-1}(a)&=& \phantom{-}\calL(a)-\frac{a}{d(1-a)}\,,
\label{relDa}\\
B^{-1}(a)&=& -\calL(a)+\frac{a}{d(1-d)(1-d-a)}\,,
\label{relBa}
\eea
with the abbreviation
\beq
\calL(a) = \frac{1}{d^2}\log\frac{(1-d)(1-a)}{1-d-a}\,.
\label{dcalL}
\eeq
Figure \ref{fig4} shows the functions $f(x)$ and $g(y)$ of
Eqs.\,(\ref{hypf})-(\ref{hypg})
with $f_{\rm c}(x)$ and $g_{\rm c}(y)$ given by
(\ref{solutionfc})-(\ref{solutiongc})
for the special case of asymmetry parameter $d=\tfrac{1}{3}$
and with $a$, $q_1$, and $q_2$
having the values that will be determined below.

At this stage the unknown parameters are $a, q_1, q_2,$ and $K$,
the last three of which have been eliminated by the differentiations
carried out above.
To determine these we have to
substitute the solutions (\ref{solutionfc}) and (\ref{solutiongc})
in the original equations (\ref{eqGyf}) and (\ref{eqGxg})
and then impose (\ref{eqzerox}) and (\ref{eqzeroy}).
\vspace{2mm}

\subsubsection{Substitution in (\ref{eqGyf}).}

Once (\ref{solutionfc}) has been substituted in (\ref{eqGyf})
and the integrals have been carried out explicitly,
we find that $\ovG_1[\xx;g(\yy)]$
becomes a linear expression in $x$.
The equality $\ovG_1[\xx;g(\yy)]=K$ then leads to
two conditions from which $q_2$ and $K$ may be solved in
terms of $a$. The result is
\beq
q_2 = \frac{A_D}{A_D-1}
\label{xq2}
\eeq
with
\beq
A_D=\frac{1}{2} - \frac{(2-a)D}{2(1-a)}\,,
\label{xAD}
\eeq
and
\beq
K = 1 - (1-q_2)\frac{2aD}{1-a}\,.
\label{xED}
\eeq
When (\ref{xq2}) together with (\ref{xAD})
is substituted in (\ref{xED}), this yields an
expression for $K$ solely in terms of $a$, the normalization constant $D$
whose $a$ dependence is given by (\ref{relDa}), and the asymmetry parameter
$d$.
\vspace{2mm}

\subsubsection{Substitution in (\ref{eqGxg}).}

Once (\ref{solutiongc}) has been substituted in (\ref{eqGxg})
and the integrals have been carried out explicitly,
we similarly find that $\ovG_2[\yy;f(\xx)]$
becomes a linear expression in $y$.
The equality $\ovG_2[\yy;f(\xx)]=-K$ then leads to
two conditions from which $q_1$ and $K$ may be solved in
terms of $a$. A third condition arises from the requirement that also for
$\yy=0$ we must have $\ovG_2[0;f(\xx)]=-K$.
It turns out, however, that this third condition is satisfied when the former
two are.
The result is
\beq
q_1 = \frac{A_B}{A_B+1}
\label{xq1}
\eeq
with
\beq
A_B = -\frac{1}{2} + \frac{B(2-2d-a)}{2(1-d)(1-d-a)}\,,
\label{xAB}
\eeq
and
\beq
K = 2d - 1 - (1-q_1)\Big[ d -
\frac{(2-d)aB}{(1-d)(1-d-a)} \Big].
\label{xEB}
\eeq
When (\ref{xq1}) together with (\ref{xAB})
is substituted in (\ref{xEB}), this yields again an
expression for $K$ solely in terms of $a$, the normalization constant $B$
whose $a$ dependence is given by (\ref{relBa}), and the asymmetry
parameter $d$.
\vspace{2mm}

\subsubsection{Combining the two substitutions.}

\begin{figure}[b]
\begin{center}
\scalebox{.45}
{\includegraphics{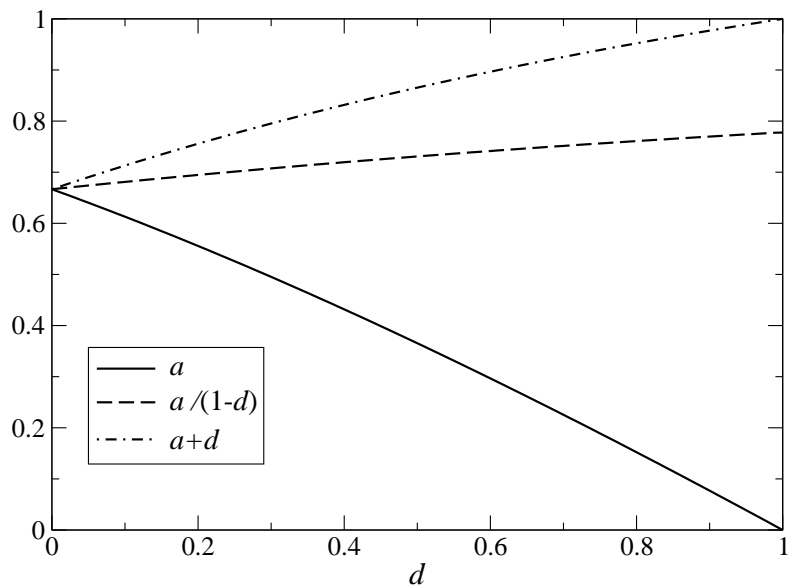}}
\end{center}
\caption{\small Interval length $a$, scaled interval length $a/(1-d)$,
and the sum $a+d$
as a function of the asymmetry parameter $d$.}
\label{fig1}
\end{figure}

\begin{figure}[hbt]
\begin{center}
\scalebox{.45}
{\includegraphics{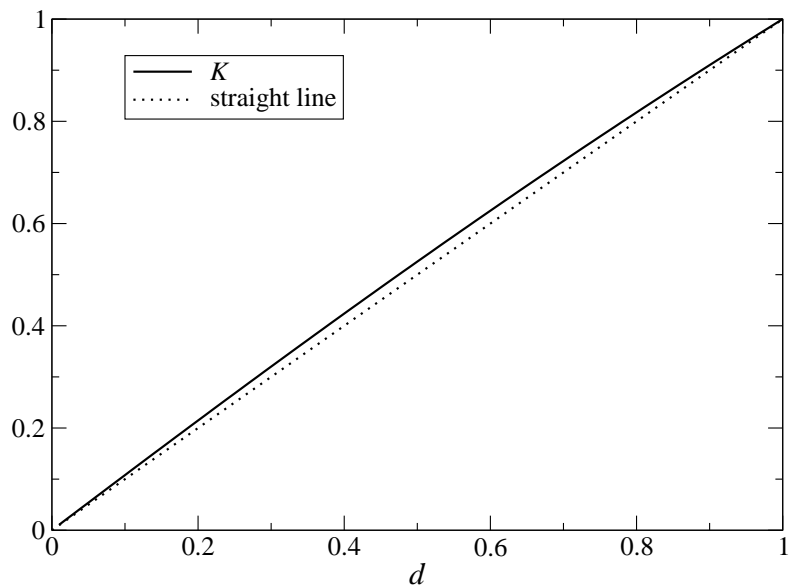}}
\end{center}
\caption{\small Expected gain $K$ of runner  \Jun\
(the stronger one) as a function of the asymmetry parameter $d$.
The dotted straight line is shown for comparison.}
\label{fig2}
\end{figure}

\begin{figure}[hbt]
\begin{center}
\scalebox{.45}
{\includegraphics{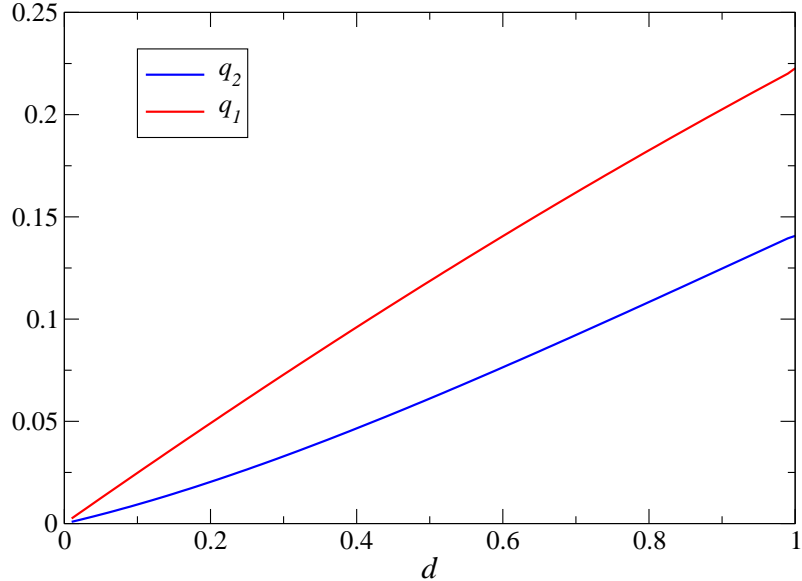}}
\end{center}
\caption{\small Coefficients $q_1$ and $q_2$ of the Dirac deltas in
Eqs.\,(\ref{hypf})
  and (\ref{hypg}) as functions of the asymmetry parameter $d$.}
\label{fig3}
\end{figure}

The interval length $a$ is fixed by the condition
that the two expressions for $K$, (\ref{xED}) and (\ref{xEB}),
be equal. After division by $BD$ this condition takes the form
\beq
\frac{2a(2-d-a)}{(1-d)(1-a)(1-d-a)} + \frac{a}{(1-d)(1-d-a)}\,\frac{1}{D}
-\frac{2}{B} - \frac{1}{BD} = 0.
\label{EXPR2}
\eeq
Upon substituting in (\ref{EXPR2}) the explicit results
(\ref{relDa})-(\ref{relBa})
for $D^{-1}$ and $B^{-1}$ we obtain a quadratic expression in
the logarithm $\calL(a)$ of Eq.\,(\ref{dcalL}) with coefficients that are
ratios of low-degree polynomials in $a$ and $d$. One may show by tedious work
that this expression can be factorized and that
(\ref{EXPR2}) reduces to
\beq
\Big[ (1-a)d\,\calL(a)-a \Big]
\Big[ (1-d-a)d\,\calL(a)-(2d+1)a-2d(d-1) \Big] =0.
\label{EXPR7}
\eeq
The first factor in (\ref{EXPR7}) does not have a zero in
  the interval $0<a<1-d$.
It thus appears that the ``physical'' solution is obtained by
setting the second factor equal to zero.
The resulting equation, slightly rewritten and exponentiated,
becomes
\beq
-\ee^{-(2d^2+1)} \,=\, -\frac{(1-d)(1-a)}{1-d-a}\,
\ee^{-\frac{(1-d)(1-a)}{1-d-a}},
\label{EXPR8}
\eeq
which is of the form $Y=X\ee^X$ with $Y=-\exp\big( -(2d^2+1) \big)$ and
$X=-(1-d)(1-a)/(1-d-a)$. The solution is $X=\calW(Y)$
where $\calW$ is the Lambert function. For negative $X$, as is our case, it has
two branches, $\calW_0$ and $\calW_{-1}$;
requiring that $0<a<1-d$ leads us to identify the
``$-1$'' branch as the ``physical'' one. Hence we have
\bea
-\frac{(1-d)(1-a)}{1-d-a} &=& \calW_{-1}\big( -\ee^{-(2d^2+1)} \big)
\nonumber\\[2mm]
&\equiv& -w\,,
\label{brutesoln}
\eea
where the second line defines $w$.
The number $w$ is positive and as $d$ varies it is in the interval
$w\in [-\calW_{-1}(-\ee^{-1}),-\calW_{-1}(-\ee^{-3})]=[1, 4.505241]$,
the lower and upper limits occurring for $d=0$ and $d=1$,
respectively.
Eq.\,(\ref{brutesoln}) now shows that the hitherto unknown
interval length $a$ is given by the elegant
expression
\beq
a =\,\frac{(1-d)(1-w)}{1-d-w}\,.
\label{xad}
\eeq
Figure \ref{fig1} shows the behavior of $a$ as a function of $d$.
It appears that for $d\to 0$ the solution of the symmetric problem with
$a=2/3$ is recovered.
Figure \ref{fig2} shows the expected gain $K$ of runner \Jun\
as a function of $d$. The values $K=0$ and $K=1$ correspond to runner \Jun\
winning with probability $\tfrac{1}{2}$ and $1$, respectively.
The increase with $d$ appears to be almost, but not completely, linear.
Figure \ref{fig3} shows the coefficients (integrated values)
$q_1$ and $q_2$ of the delta peaks
in the strategies, as a function of the asymmetry parameter $d$.
For $d\to 0$ the amplitudes of the delta functions vanish and
the solution of the symmetric problem is recovered.
Figure \ref{fig4} shows, for the special value $d=\tfrac{1}{3}$,
the strategies $f(\xx)$ and $g(\yy)$
that are the main result of this report.
We have plotted them as functions of the energies $\epsilon_1=x+d$ and
$\epsilon_2=y$.
Each strategy consists of a
continuous part and a Dirac delta. The delta functions have been presented
as vertical bars of height equal to 100 times their integrated value
in the corresponding point, {\it viz.}
$x=0$ for $f(x)$
and $y=-\frac{1}{3}$ for $g(y)$.

In summary, we found a pair of functions $f(x)$ and $g(y)$
given by Eqs.\,(\ref{hypf}) and (\ref{hypg}),
(\ref{solutionfc})-(\ref{dcalL}) for $f_{\rm c}(x)$ and $g_{\rm c}(y)$,
Eqs.\,(\ref{xq2})-(\ref{xAD}) and (\ref{xq1})-(\ref{xAB})
for the parameters $q_2$ and $q_1$, respectively, in which finally the
interval length $a$ is fixed by
Eqs.\,(\ref{brutesoln})-(\ref{xad}).


\subsection{Completing the proof}
\label{sec:completing}

In order to complete the proof that this pair is a Nash equilibrium, we have
to show that neither player can improve his gain by choosing a strategy
outside of the support  of the functions
($f(x)$ for \Jun\ and $g(y)$ for \Jdeux) obtained above.
That is, we must show that
\bea
\ovG_1[\xx;g(\yy)] \leq \ \ K, \qquad && \xx\notin\mbox{supp } f,
\label{expectx}\\[2mm]
\ovG_2[\yy;f(\xx)] \leq -K,    \qquad && \yy\notin\mbox{supp } g,
\label{expecty}
\eea
which complement Eqs.\,(\ref{eqzerox})-(\ref{eqzeroy}).
We will prove Eqs.\,(\ref{expectx})-(\ref{expecty}) below.

Let us first consider runner \Jun\ and suppose that he
chooses a strategy $x\in(a,1]$.
Then his expected gain is
\bea
\ovG_1[\xx;g(\yy)] &=& q_2 G_1(x,0)
                       + (1-q_2)\int_d^{a+d}\dd y\,G_1(x,y)g_{\rm c}(y)
\nonumber\\[2mm]
&=& q_2(1-2x) + (1-q_2)\int_d^{a+d}\dd y\,(1-2x+xy)g_{\rm c}(y),
\label{Bone}
\eea
where the expression for $G_1(x,y)$ is a consequence of the fact that
here $x>a$ and $y\leq a+d$, whence $x+d>y$.
Figure \ref{figxysquarelarge} is helpful to visualize these various
inequalities.
Expression (\ref{Bone}) is clearly strictly decreasing with $x$ and therefore
\beq
\ovG_1[x;g(y)]<\ovG_1[a;g(y)] = K \qquad \mbox{ for } x\in(a,1],
\label{Bineqone}
\eeq
which means that (\ref{expectx}) is satisfied.

Let us next consider runner \Jdeux.   Three cases may occur.

{\it Case 1.} If \Jdeux\ chooses a strategy $y\in(a+d,1]$,
his expected gain is
\bea
\ovG_2[\yy;f(\xx)] &=& -q_1 G_1(0,y) - (1-q_1)\int_0^{a}\dd x\,G_1(x,y)
                        f_{\rm c}(x)
\nonumber\\[2mm]
&=& q_1(1-2y) + (1-q_1)\int_0^{a}\dd x\,(1-2y+xy)f_{\rm c}(x),
\label{Btwo}
\eea
where the expression for $G_1(x,y)$ is a consequence of the fact that
here $x\leq a$ and $y>a+d$, whence $x+d<y$.
Expression (\ref{Btwo}) is clearly
strictly decreasing with $y$ and therefore
\beq
\ovG_2[y;f(x)]<\ovG_2[a+d;f(x)] = -K \qquad \mbox{ for } y\in(a+d,1].
\label{Bineqtwo}
\eeq
This means that (\ref{expecty}) is satisfied for $y\in(a+d,1]$.

{\it Case 2.} If \Jdeux\ chooses a strategy
$y\in(0,d)$, his expected gain is
\bea
\ovG_2[\yy;f(\xx)] &=& -q_1 G_1(0,y) - (1-q_1)\int_0^{a}\dd x\,G_1(x,y)
                        f_{\rm c}(x)
\nonumber\\[2mm]
&=& -q_1 - (1-q_1)\int_0^{a}\dd x\,(1-2x+xy)f_{\rm c}(x),
\label{Bthree}
\eea
where the expression for $G_1(x,y)$ is a consequence of the fact that
here $x \geq 0 $ and $y<d$, whence $x+d>y$.
Expression (\ref{Bthree}) is clearly strictly decreasing with $y$ and
therefore
\beq
\ovG_2[y;f(x)]<\ovG_2[0;f(x)] = -K \qquad \mbox{ for } y\in(0,d),
\label{Bineqthree}
\eeq
and so (\ref{expecty}) is also satisfied for $y\in(0,d)$.

{\it Case 3.} 
If \Jdeux\ chooses the strategy $y=d$, his expected gain is
still given by the first line of (\ref{Bthree}), but now
with $G_1(0,d)=\gamma(0)$
since $(0,d)$ is on the line of discontinuity. Hence
\beq
\ovG_2[d;f(x)] = -q_1 (d+\gamma(0))
- (1-q_1)\int_0^{a}\dd x\,G_1(x,d)f_{\rm c}(x).
\label{Bfour}
\eeq
We now consider Eq.\,(\ref{eqGyf}), which holds for $y>d$.
When taking the limit $y\to d$ and using that $\lim_{y\to d}G(0,y)=-1+2d$,
we obtain
\beq
\lim_{y\to d}\ovG_2[y;f(x)] =
q_1(1-2d) - (1-q_1)\int_0^a\dd x\,G_1(x,d)f_{\rm c}(x) = -K.
\label{Bfive}
\eeq
By comparing (\ref{Bfour}) and (\ref{Bfive}) we see that
\beq
\ovG_2[d;f(x)] = -K -(1-d+\gamma(0))q_1 \leq -K,
\label{Bineq4}
\eeq
where in the last step we used that
$0 \leq 1-d-\gamma(0)=(1+\gamma_0)(1-d)$.
So (\ref{expecty}) is also satisfied for $y=d$.

Note that for $\gamma_0=-1$ the inequality in Eq.\,(\ref{Bfive})
holds with the
equality sign and we may in that case include the boundary point
$y=d$ in the support of $g_{\rm c}(y)$.

The combined results of these three cases
demonstrate that neither player can improve his gain
by employing a strategy that is outside the support that we found for him, and
this means our solution is a Nash equilibrium.


\subsection{Limits $d\to 0$ and $d\to 1$}

We observed numerically in Figures \ref{fig1}, \ref{fig2}, and \ref{fig3}
that in the limit $d\to 0$, we recover the symmetric results.
We shall now find analytically how this limit case is approached.
In order to recover the Nash equilibrium of the symmetric case we have to let
$d\to 0$ in Eq.\,(\ref{xad}).
The small $d$ behavior of $w$ requires
the expansion of the Lambert function about its branch point
\footnote{The full expansion is given by Corless
{\it et al.}\,\cite{corless1996}, Eq.\,(4.22).},
 which leads to
\beq
w = 1 + 2d + \tfrac{4}{3}d^2 + \frac{11}{9}d^3 + {\cal O}(d^4).
\label{ewsmalld}
\eeq
When substituted in (\ref{xad}) we find that
\beq
a = \tfrac{2}{3}\Big( 1-\tfrac{7}{9}d + {\cal O}(d^2) \Big),
\label{easmalld}
\eeq
which for $d=0$ reproduces the value found in section \ref{secsymm}.
To find the
corresponding small $d$ expansions of $q_1$ and $q_2$ ,
the most convenient way of doing is to first expand (\ref{relDa}) and
(\ref{relBa})
for small $d$ at fixed $a$, which leads to a cancelation
of the terms proportional to $d^{-1}$, and to then insert (\ref{easmalld}).
The result is that
\beq
D^{-1}(a) = 4\Big( 1-\frac{4}{3}d+{\cal{O}}(d^2) \Big), \qquad
B^{-1}(a) = 4\Big( 1+\frac{5}{6}d+{\cal{O}}(d^2) \Big).
\label{eDBsmalld}
\eeq
Upon combining this expansion of $D^{-1}(a)$ with
(\ref{xAD}) and (\ref{xq2}), and the one of $B^{-1}(a)$ with
(\ref{xAB}) and (\ref{xq1}) we find
a small-asymmetry behavior of the delta peak strengths,
\beq
q_2 = \tfrac{1}{12}d + {\cal O}(d^2), \qquad
q_1 = \tfrac{1}{4}d + {\cal O}(d^2).
\label{eq12smalld}
\eeq
When the expansions of $q_2$ and $D$ are substituted in (\ref{xED}), or,
alternatively,
those of $q_1$ and $B$ in (\ref{xEB}), we find that in the
small $d$ limit the expected gain $K$ of runner \Jun\ is given by
\beq
K = \frac{13}{12}d + {\cal O}(d^2),
\label{esmalldK}
\eeq
which explains the very small deviation of $K$ from
a straight line in Figure \ref{fig2}.

In the limit $d\to 1$ the interval $[0,1-d]$ tends to zero and so necessarily
$a\to 0$.
However, we find that the limit
\beq
\lim_{d\to 1}\frac{a}{1-d} = 1+1/{\calW_{-1}(-\ee^{-3})} = 0.778036
\label{limdone}
\eeq
corresponds to a finite value, as can also be seen in Fig.~\ref{fig1}.
It is clear from this limit behavior
that $a$ stays away from the end points of the interval $[0,1-d]$.


\section{Conclusion}
\label{sec:conclusion}

We have introduced a model which describes the competition between two athletes
 who have a difference of average strength described by a parameter $d$.

The athletes ($R_1$ and $R_2$)
interact through their choice of a strategy ($x$ and $y$, respectively),
which is essentially the amount of energy they invest in the competition.
The choice of a strategy is based on their knowledge of each other's
average performance and their evaluation of the danger of
exhaustion, which increases with the amount of invested energy.

We have formulated this problem as a zero-sum game.
The symmetric version of the problem, $d=0$, is well-known in the game theory
literature and has a Nash equilibrium in which both athletes optimize
their chances for victory by adopting mixed strategies.

We have studied here the asymmetric game, $d>0$.
Mathematically, this problem is in the class of ``discontinuous games''
and the existence of a Nash equilibrium is not guaranteed in advance.
We have first
shown that there is no Nash equilibrium in pure strategies.
We have then demonstrated by explicit construction
that the problem has a mixed strategy Nash equilibrium for arbitrary $0<d<1$.
For each athlete the best mixed strategy appears to be the sum of a continuous
distribution and a Dirac delta peak.
It is remarkable that this problem is analytically tractable.
We have not addressed the question of the uniqueness of the solution that we have
found; but our numerical work leads us to believe that it is unique.

Our solution shows in particular that, in the case where
the energy distributions have an overlap ($0<d<1$), the weaker athlete
has a nonvanishing chance to win
against the stronger one, as expected, and provides him with the best
strategy.
 In particular, according to our results, the weaker athlete should sometimes choose a
very cautious strategy (corresponding to the Dirac peak at $y=0$),
so as to benefit of any misfortune (injuries, false start, etc) of the stronger runner.
Real sport 
competitions~\cite{moussambani2000} provide examples, albeit rare ones,
of such events.
In our model, this effect is overemphasized due to the sharp distinction
between optimal running and exhaustion.
In real competitions, there is a continuous transition between these two states,
that could be included in further modeling.

In this paper, we have considered that both runners have the same
knowledge about their own energy distribution and the one of
their competitor.
In practice, a runner has a better knowledge of his own.
However, due to the fact that
we have a sharp transition between exhaustion and optimal
running,
taking a narrower distribution for one's own
energy distribution would not deeply change the results.

In future work, several directions of research could be explored.

As we discussed in the paper, what we call {\em energy} is rather
an aggregation of several parameters. These parameters could
be explicitly considered.
Then it would be possible for example that one athlete would be stronger in terms
of anaerobic energy but weaker in terms of VO2
(in any case, we could still define the stronger athlete as the one
having the shortest time when running alone).
However, it will probably be quite difficult to handle analytically
a model with more energy variables.

In the present model, once a runner has chosen his strategy
through the choice of an energy on which he will count,
he runs as if he were alone.
An important step would be to consider a race time that itself
depends on both athletes, allowing to include some interaction
effect during the race.

As mentioned earlier, though we have presented our model in the framework
of athletics' running, other types of competitions could be considered.
For some of them (for example shot put or long jumps), the time variable
$T(\EE_i)$ should be replaced by a length.
However this would not change the results that we have obtained
in the very general framework proposed in this paper.
The model should become more sport dependent when
describing in a more refined way how $T$ depends on the
different physiological
characteristics of the athlete, or when considering
interactions between the athletes during the competition itself.

Implicitly, in our choice for the expression of the payoff,
we assume that each athlete wants to minimize the number of
competitions that he loses. This may not necessarily be the case.
For example, a runner could prefer to have a few brilliant victories
(meaning that he would arrive much in advance compared to the other athletes),
even if this means losing more races. Then the gain should be written
in a way that reflects the goal that the athlete is trying to achieve.

Beyond these variants around our model, which are all static models,
another perspective would be to consider a differential game~\cite{bressan2011},
{\it i.e.,} a game in which some time evolution is included, to describe
how athletes adapt their strategy throughout the race.
Though analytical solutions will then be out of reach,
it could be very instructive to explore various types of strategies
in this framework.


\section*{Acknowledgments}

We would like to thank Rida Laraki for pointing out relevant literature to us
and for valuable remarks.

C.A.-R. acknowledges support for traveling from CNRS (D\'efi S2C3 - AAP2017, project GAMEPED).


\appendix

\section{\,\,\,Proof of Eq.\,(\ref{xsupinf})}
\label{sec:AppendixA}

We wish to calculate
$\sup_{x} \inf_{y} G_1(x,y)$ and $\inf_{y} \sup_{x} G_1(x,y)$.
We consider here the latter.
To calculate $\sup_{x} G_1(x,y)$ we distinguish two different
$y$ intervals, namely $0\leq y<d$ and $d\leq y\leq 1$.
For  $0\leq y<d$ we have, using expression (\ref{game_asym}) for $G_1(x,y)$,
\beq
\sup_x G_1(x,y) = \sup_x(1-2x+xy) =1, \qquad 0\leq y<d,
\label{supx1}
\eeq
there being a maximum for $x=x^*(y) = 0$.
For   $d\leq y\leq 1$ we begin by calculating $\sup_x$ separately for
$0\leq x < y-d$,\, $x=y-d$, and $y-d<x\leq 1$, and then take the maximum of
the three results. This yields, again with the aid of (\ref{game_asym}),
\bea
\sup_x G_1(x,y) &=& \max\left[ \sup_{0\leq x<y-d}(-1+2y-xy),\, d+\gamma(y-d),\,
                            \sup_{y-d<x\leq 1}(1-2x+xy) \right]
\nonumber\\[2mm]
&=& \max[ -1+2y,\, d+\gamma(y-d),\, 1-(2-y)(y-d) ],
\nonumber\\[2mm]
&& \qquad \qquad d\leq y\leq 1.
\label{supx2}
\eea
We note that for $y=d$ the interval $0\leq x<y-d$ that appears above is empty and the corresponding argument of the maximum operator is absent.
Of the three arguments of the maximum in the last line,
the first one has been
obtained for $x^*(y)=0$; the second and third ones both for
$x^*(y)=y-d$, which corresponds to the line of discontinuity.
The third argument does not exist as a maximum but is a supremum.

It may be verified with the aid of expression (\ref{dgamma})
that $d+\gamma(y-d)$ is never larger than either
$-1+2y$ or $1-(2-y)(y-d)$. Of these, the former increases and the latter decreases with $y$ on $[d,1]$. Their point of intersection occurs at $y=y^*$ given by
\beq
y^* = \frac{1}{2}\left[\,4+d-\sqrt{8+d^2}\,\right].
\label{xystar}
\eeq
Upon combining the results of the two $y$ intervals we therefore have
\beq
\sup_x G_1(x,y) = \left\{
\begin{array}{ll}
\phantom{-}1 & 0\leq y \leq d, \\[2mm]
\phantom{-}1-(2-y)(y-d) \quad & d\leq y \leq y^*, \\[2mm]
-1+2y & y^* \leq y \leq 1.
\end{array}
\right.
\label{supxG}
\eeq
This shows that $\sup_x G_1(x,y)$ reaches its minimum as a function of $y$
for $y=y^*$.
We therefore have
\bea
\infy\sup_x G_1(x,y) &=& -1 + 2y^* \nonumber\\[2mm]
&=&\,\, d+3-\sqrt{8+d^2},
\label{infsupG}
\eea
the $\inf_y\sup_x$ being realized for $(x,y)=(x^*(y^*),y^*)=(y^*-d,y^*)$,
which is a point on the line of discontinuity.
By similar methods but that we will not detail here
one derives the counterpart
\beq
\sup_x\infy G_1(x,y) =  d-3+\sqrt{8+d^2}.
\label{supinfG}
\eeq
Eqs.\,(\ref{infsupG}) and (\ref{supinfG}) are the results stated in Eq.\,(\ref{xsupinf}).

\section*{References}

\appendix

\end{document}